\documentclass[runningheads]{llncs}
\usepackage{graphicx}
\usepackage{bbding}
\usepackage{graphicx}
\usepackage{float}
\usepackage{subfigure}
\usepackage{booktabs}
\usepackage{makecell}
\usepackage{multirow}
\usepackage{caption}

\def\ie{\emph{i.e.}, }

\def\etal{\emph{et al.}}

\begin{document}
\title{Boosting Neural Video Representation via Online Structural Reparameterization}
\titlerunning{Boosting NVR via Online Structural Reparameterization}

\author{Ziyi Li\inst{1} \and
Qingyu Mao\inst{1} \and
Shuai Liu\inst{2,3} \and
Qilei Li\inst{5} \and
Fanyang Meng\inst{4} \and
Yongsheng Liang\inst{1,3(}\Envelope\inst{)}
}
\authorrunning{Z. Li et al.}
\institute{College of Electronics and Information Engineering, Shenzhen University, Shenzhen, China \and
College of Applied Technology, Shenzhen University, Shenzhen, China \and
College of Big Data and Internet, Shenzhen Technology University, Shenzhen, China \and
Research Center of Networks and Communications, Peng Cheng Laboratory, Shenzhen, China \and
Faculty of Artificial Intelligence in Education, Central China Normal University, Wuhan, China \\
\email{liangys@szu.edu.cn} }
\maketitle              
\begin{abstract}
Neural Video Representation~(NVR) is a promising paradigm for video compression, showing great potential in improving video storage and transmission efficiency.
While recent advances have made efforts in architectural refinements to improve representational capability, these methods typically involve complex designs, which may incur increased computational overhead and lack the flexibility to integrate into other frameworks.
Moreover, the inherent limitation in model capacity restricts the expressiveness of NVR networks, resulting in a performance bottleneck.
To overcome these limitations, we propose Online-RepNeRV, a NVR framework based on online structural reparameterization.
Specifically, we propose a universal reparameterization block named ERB, which incorporates multiple parallel convolutional paths to enhance the model capacity.
To mitigate the overhead, an online reparameterization strategy is adopted to dynamically fuse the parameters during training, and the multi-branch structure is equivalently converted into a single-branch structure after training.
As a result, the additional computational and parameter complexity is confined to the encoding stage, without affecting the decoding efficiency.
Extensive experiments on mainstream video datasets demonstrate that our method achieves an average PSNR gain of 0.37-2.7 dB over baseline methods, while maintaining comparable training time and decoding speed.

\end{abstract}

\section{Introduction}
Neural Video Representation~(NVR) has attracted significant attention for its effectiveness in modeling video data.
Specifically, NVR represents videos as continuous functions parameterized by neural networks, which map input coordinates to their corresponding values.
This paradigm has enabled a wide range of applications in various video-related tasks, such as compression~\cite{zhang2021vcompress,gao2025pnvc}, inpainting~\cite{ji2025vinpainting,kim2024snerv}, and interpolation~\cite{lee2023ffnerv,yan2024dsnerv}.

\begin{figure}[t]
\centering
\includegraphics[width=\textwidth]{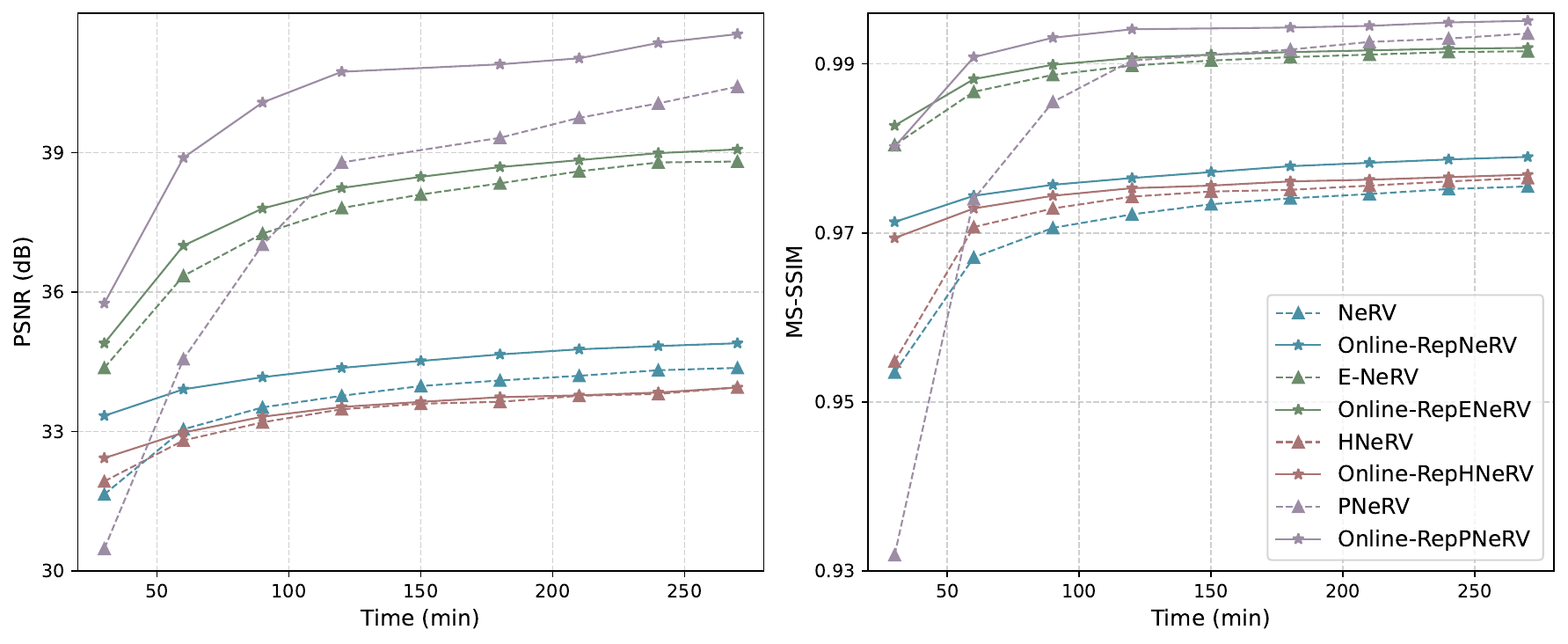}
\caption{Performance curves of four baseline methods and their Online-Rep counterparts under the same training time. Online-Rep achieves higher PSNR and MS-SSIM in the early training stages and maintains a performance advantage.} 
\label{fig1}
\end{figure}

In video compression, NeRV~\cite{chen2021nerv} introduces an innovative image-wise representation framework that leverages MLP and convolutional layers to map temporal index to its corresponding frame.
This approach formulates the encoding process as training a neural network to overfit the input video, with the resulting network parameters serving as a compact representation of the video.
These parameters are subsequently compressed and transmitted to the decoder for reconstruction.
Although NeRV exhibits high efficiency, its representational capacity is constrained by its relatively simple structure.
Consequently, a series of subsequent studies~\cite{li2022e-nerv,zhao2023dnerv,kwan2024hinerv,tarchouli2024res-nerv,zhao2024pnerv} have investigated various architectural enhancements aimed at improving the representational capacity and reconstruction quality of NeRV-based models.
For instance, HiNeRV~\cite{kwan2024hinerv} combines lightweight layers with hierarchical positional encodings to build a deep and wide network architecture,
and PNeRV~\cite{zhao2024pnerv} employs a pyramid structure to capture information from adjacent frames.
While effective, these methods typically incorporate complex components, such as transformer-based modules~\cite{li2022e-nerv} or hierarchical designs~\cite{zhao2024pnerv}, which substantially increase the computational burden during decoding.
In addition, these components are tightly coupled with their corresponding backbone networks, preventing them from being flexibly integrated or used interchangeably.

Beyond architectural enhancements, empirical observations indicate that the representational power of NVR models is fundamentally constrained by their limited capacity (model complexity)~\cite{chen2021nerv}.
Specifically, in NVR frameworks, the encoding is an overfitting process, implying that prolonging the training procedure typically results in improved reconstruction performance.
However, as illustrated in Fig.~\ref{fig1}, this performance improvement eventually plateaus as training progresses, suggesting the presence of an inherent bottleneck associated with model capacity.
In other words, beyond a certain point, merely extending training time fails to yield further meaningful improvements in reconstruction quality due to the fundamental limitations imposed by model size and complexity.
Motivated by these challenges, we raise the following central question: \textit{Is it possible to effectively scale up the model's capacity to achieve enhanced performance without incurring additional computational costs during decoding?}

To address this issue, we explore structural reparameterization, a technique that has demonstrated significant success in various computer vision tasks, including classification and super-resolution. 
Structural reparameterization decouples the training-time and inference-time architectures, thereby enabling increased model capacity during training without incurring additional computational overhead during decoding.
Motivated by these advantages, we propose Online-RepNeRV, a novel NVR framework based on online structural reparameterization. Specifically, we first empirically investigate conventional reparameterization blocks within the NVR setting and find that their performance improvements are limited.
To resolve this limitation, we introduce an Enhanced Reparameterization Block (ERB), a universal module explicitly tailored for NeRV-based architectures.
ERB employs a multi-branch structure comprising parallel convolutional kernels with diverse kernel sizes, enabling the extraction of richer and more comprehensive features across multiple receptive fields.
To mitigate the increased computational burden typically associated with multi-branch designs, we further propose an online reparameterization strategy that dynamically fuses branch parameters during training. 
This online fusion strategy significantly reduces training overhead while fully preserving the representational benefits of the expanded structure. 
After training, the multi-branch ERB can be equivalently merged into a standard convolutional layer, ensuring that transmission and decoding efficiency remain unaffected.
In summary, our contribution are three-fold:
\begin{itemize}
\item[$\bullet$] We present the first systematic exploration of structural reparameterization within NVR, and propose Online-RepNeRV, a plug-and-play framework for enhancing representational capacity without increasing decoding complexity.
\item[$\bullet$] We design an Enhanced Reparameterization Block (ERB), which integrates multi-scale convolutional branches and employs an online parameter fusion strategy, effectively balancing model expressiveness and training efficiency.
\item[$\bullet$] Extensive experiments validate that Online-RepNeRV can be flexibly integrated into various NVR architectures and outperforms existing baselines, highlighting its broad applicability and effectiveness.
\end{itemize}

\section{Related Work}
\textbf{Neural Video Representation} has emerged as a promising paradigm for efficient video modeling and compression.
One representative work, NeRV~\cite{chen2021nerv}, introduces an image-wise implicit representation that takes the frame index as input and directly outputs the entire frame.
Based on NeRV, a series of methods~\cite{li2022e-nerv,zhao2023dnerv,kwan2024hinerv,tarchouli2024res-nerv,zhao2024pnerv,chen2022cnerv,chen2023hnerv} have focused on enhancing video representation and reconstruction quality.
E-NeRV~\cite{li2022e-nerv} disentangles spatial-temporal contexts and encodes them using transformer and MLP respectively, significantly improved encoding speed while maintaining high reconstruction quality.
By employing ConvNeXt blocks~\cite{liu2022convnet} to generate tiny content-adaptive embeddings before a learnable decoder, HNeRV~\cite{chen2023hnerv} combines the advantages of explicit and implicit representations, offering better generalization and compression performance.
Zhao \etal~\cite{zhao2023dnerv} proposes a differential neural representation (DNeRV) that encodes the differences between adjacent frames and fuses them with content feature stream, effectively solving the temporal inconsistency problem during reconstruction.
Subsequently, PNeRV~\cite{zhao2024pnerv} further addresses the problem of spatial inconsistency by constructing a pyramid information interaction structure, which fuses content and differential feature representation across multiple levels.
By integrating residual blocks~\cite{he2016resblock} into NeRV, Res-NeRV~\cite{tarchouli2024res-nerv} enhances the reconstruction of both fine-grained textures and high-level features.
However, these methods typically involve complex designs that may introduce computational overhead during inference and lack flexibility for integration into other frameworks.
In contrast, our proposed Online-RepNeRV addresses these limitations by employing a plug-and-play online structural reparameterization strategy, significantly enhancing model capacity without additional decoding overhead and allowing flexible integration into various NVR architectures.

\noindent\textbf{Structural reparameterization} techniques are widely adopted across various domains, including object detection~\cite{ding2019acnet,ding2021dbb,ding2021repvgg}, super resolution~\cite{zhang2021ecbsr,deng2023reprfn,liu2023telnet}, and neural architecture search (NAS)~\cite{zhang2023repnas,yu2023efficient}.
ACNet~\cite{ding2019acnet} proposes using 1$\times$3 and 3$\times$1 asymmetric convolutions to reinforce the central skeleton of standard 3$\times$3 convolution.
DBB~\cite{ding2021dbb} enhances a single convolution through the integration of diverse branches with varying types, including multi-scale convolutions, sequential convolutions and average pooling.
RepVGG~\cite{ding2021repvgg} improves the performance of the conventional VGG architecture by constructing a 3$\times$3 convolution with parallel identity mappings and 1$\times$1 convolutions.
Zhang \etal~\cite{zhang2021ecbsr} propose an Edge-oriented Convolution Block (ECB), which contains standard convolutions and a set of learnable edge-aware filters. 
This design effectively captures edge and texture information, leading to enhanced reconstruction quality in super-resolution tasks.
RepRFN~\cite{deng2023reprfn} incorporates all the aforementioned branches to effectively capture multi-scale information and high-frequency edge features.
Although these approaches have demonstrated their effectiveness in vision-related tasks, our experiments reveal that their direct application to NVR yields limited improvements.
In this work, we propose the Enhanced reparameterization Block (ERB) for NVR based on structural reparameterization.
It can serve as a drop-in replacement for standard 3$\times$3 convolutions in NeRV-based architectures, enabling richer model capacity during training.

\section{Method}
\subsection{Motivation and Overall architecture}
\label{Motivation}
As discussed in the introduction, incorporating complex network components tends to increase computational overhead.
Meanwhile, a straightforward solution to address limited model capacity is to increase the number of parameters.
However, any increase in either computational or parameter complexity inevitably raises the inference cost, which is undesirable in video compression scenarios.

To address this challenge, we aim to develop a solution that introduces additional complexity only during training, while keeping the inference structure unchanged.
Structural reparameterization employs more expressive architectures during training and deploys more efficient structures at inference, providing a natural way to separate training complexity from inference cost.
Motivated by the above considerations, we investigate the application of structural reparameterization in NVR, aiming to optimize the trade-off between model expressiveness and inference efficiency.

\begin{figure}[t]
\includegraphics[width=\textwidth]{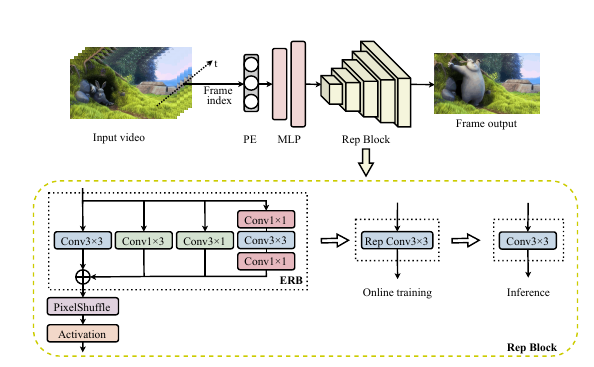}
\caption{Overall architecture of Online-RepNeRV. Taking the frame index as input, the model utilizes an MLP and multiple rep blocks to reconstruct the frame. Each rep block incorporates a multi-branch ERB, whose parameters are merged into a single set during online training, and the structure is converted into a single-branch form during inference.} \label{fig2}
\end{figure}

\begin{figure}[t]
\centering
\includegraphics[width=8cm]{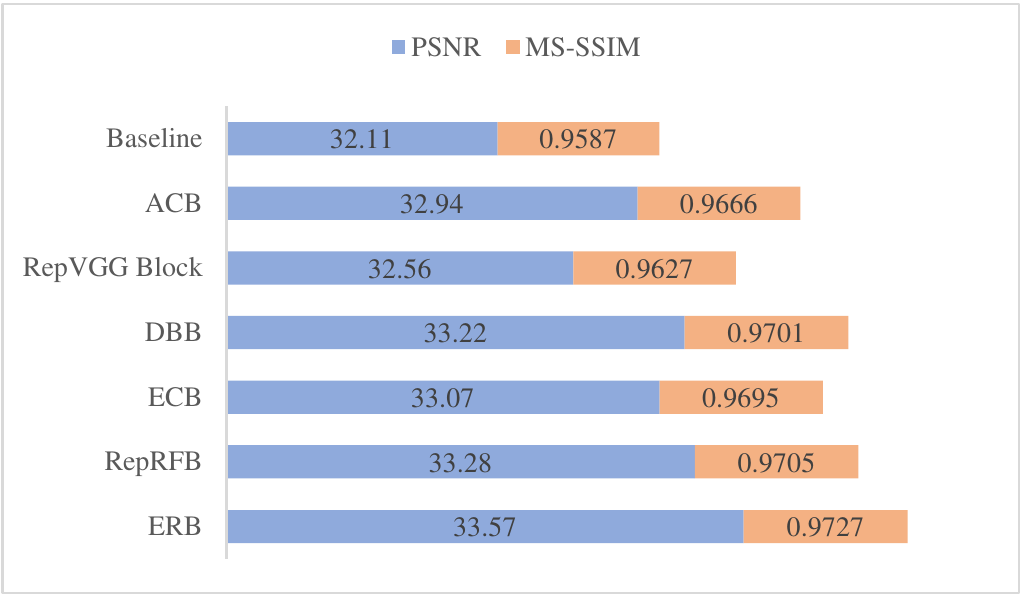}
\caption{Performance of different reparameterized blocks on NVR.} \label{fig3}
\end{figure}

We initially propose a reparameterized architecture Rep-NeRV for NeRV-style frameworks.
As depicted in the upper portion of Fig.~\ref{fig2}, Rep-NeRV incorporates positional encoding, an MLP, and multiple decoder blocks (referred to as rep blocks).
Each decoder block consists of reparameterized unit, a PixelShuffle~\cite{shi2016pixelshuffle} operation for upsampling, and an activation function.
During the training phase, structural reparameterization typically adopts multi-branch structures composed of various operations, such as convolution, pooling, and identity mapping.
To explore the feasibility of these designs in our task, we begin by revisiting classical reparameterized blocks originally developed for other vision tasks and integrate them into the rep blocks of Rep-NeRV.
Specifically, we evaluate the performance of ACB~\cite{ding2019acnet}, RepVGG Block~\cite{ding2021repvgg}, DBB~\cite{ding2021dbb}, ECB~\cite{zhang2021ecbsr}, and RepRFB~\cite{deng2023reprfn}, with the results presented in Fig.~\ref{fig3}.
Although these blocks exhibit varying degrees of effectiveness, our experiments indicate that they are not the optimal configurations in the NVR setting.
In particular, we observe that simply adding the number of branches does not guarantee better performance.
For instance, the combination of 1$\times$3, 3$\times$1 and 1$\times$1 is less effective than 1$\times$3, 3$\times$1 alone, indicating that unnecessary branches may introduce redundancy and interfere with feature learning.
By contrast, the impact of cascaded 1$\times$1-3$\times$3 convolution is consistently positive, likely due to its ability to jointly capture local and contextual information in an efficient manner.
See Section~\ref{Ablation Study} for additional ablation results.

These results highlight the importance of careful branch design in effective reparameterization and motivate the development of our proposed Enhanced Reparameterization Block~(ERB).
Building upon this, we propose the Online-RepNeRV framework, which integrates ERB and online reparameterization into Rep-NeRV.

\subsection{Enhanced Reparameterization Block}
The ERB is composed of four branches, summarized in following three types.

\textbf{Vanilla 3$\times$3 convolution} To preserve the feature extraction capability of the decoder block, we retain the 3$\times$3 convolution, which serves as the core component for capturing spatial correlations in local regions.
The output of this branch is denoted as:
\begin{equation}
F_v = K_v * X + B_v
\label{1}
\end{equation}
where F, X, K, B represent the output feature, input feature, convolution weight, and bias, respectively. The symbol * denotes the convolution operator.

\textbf{Asymmetric convolutions} As demonstrated in~\cite{ding2019acnet}, the weight distribution of square convolution kernel is uneven across different positions, with a greater concentration around the central cross region.
Therefore, we adopt 1$\times$3 and 3$\times$1 convolution to strengthen the cross skeleton of the 3$\times$3 convolution.
The output is indicated as:
\begin{equation}
F_{asy} = (K_{asy1} * X + B_{asy1}) + (K_{asy2} * X + B_{asy2})
\label{2}
\end{equation}

\textbf{Sequential convolution} Although the 1$\times$1-3$\times$3 sequential convolution has shown moderate effectiveness, we enhance feature representation by stacking an additional 1$\times$1 convolution, while maintaining equivalence to a single 3$\times$3 convolution.
This extension introduces richer nonlinear transformations and further improves the model's learning capability.
The output generated by this branch is represented as:
\begin{equation}
F_{s} = K_{s3} * (K_{s2} * (K_{s1} * X + B_{s1}) + B_{s2}) + B_{s3}
\label{3}
\end{equation}

The output of the ERB is formed by:
\begin{equation}
F = F_v + F_{asy} + F_s
\label{4}
\end{equation}

\subsection{Online Reparameterization for Efficient Training}
While multi-branch structures improve representational capacity during training, they inevitably increase computational overhead and memory consumption due to the additional parameters and operations introduced.
This can significantly slow down the training process, especially for large-scale models or datasets.

To mitigate this effect, inspired by ~\cite{hu2022online}, we adopt an online reparameterization strategy. 
Instead of deferring the merging of branches to the post-training phase, this approach separates the merging into two stages: parameters fusion occurs dynamically during training and structural fusion is performed after training.
For ERB, the parameters of its four branches are merged into a single set of parameters. The process is described as follows:

\textbf{Integration of asymmetric convolutions} For convolutions whose kernel size is smaller than the target size, we align the spatial centers with zero padding. With \{$K_{asy1}, B_{asy1}$\} and \{$K_{asy2}, B_{asy2}$\}, the 1$\times$3 and 3$\times$1 branch parameters can be merged into a single set for a 3$\times$3 convolution:
\begin{equation}
K^{'}_{asy}=pad(K_{asy1})+pad(K_{asy2})\label{5}
\end{equation}
\begin{equation}
B^{'}_{asy}=B_{asy1}+B_{asy2}\label{6}
\end{equation}

\textbf{Combination of sequential convolution} First, the parameters of the 1$\times$1 and 3$\times$3 convolutions are merged, where the 1$\times$1 convolution performs a linear combination of channels and can be integrated into the 3$\times$3 convolution via convolution operation. With \{$K_{s1}, B_{s1}, K_{s2}, B_{s2}$\}, we have:
\begin{equation}
K_{tmp} = K_{s2} * trans(K_{s1})\label{7}
\end{equation}
\begin{equation}
B_{tmp} = B_{s2} + B_{s1} * K_{s2}\label{8}
\end{equation}
where trans($K_{s1}$) denotes the tensor obtained by channel transpose of $K_{s1}$.

Next, the second 1$\times$1 convolution is incorporated into the fused result by matrix multiplication. With \{$K_{s3}, B_{s3}$\}, the final parameters are derived as follows:
\begin{equation}
K^{'}_{s} = expand(K_{s3}) \times K_{tmp}\label{9}
\end{equation}
\begin{equation}
B^{'}_{s} = B_{s3} + B_{tmp} * K_{s3}\label{10}
\end{equation}
where expand($K_{s3}$) refers to broadcasting $K_{s3}$ into a 3$\times$3 form.

\textbf{Integration of multi-branch} Leveraging the linearity of convolution, the parameters of multiple branches can be merged into a unified set. The resulting weight and bias are given by:
\begin{equation}
K^{'} = K_v + K^{'}_{asy} + K^{'}_{s}\label{11}
\end{equation}
\begin{equation}
B^{'} = B_v + B^{'}_{asy} + B^{'}_{s}\label{12}
\end{equation}

During the forward pass, the merged parameters $K^{'}$ and $B^{'}$ are used for computation, while the gradient back-propagation still traces back to the individual branches.
This approach not only speeds up training but also preserves the feature representation capability of the multi-branch structure.

Upon completion of training, the multi-branch structure is converted into a single $3\times$3 convolution.
This transformation ensures functional equivalence between training and inference network, while maintaining the inference architectures unchanged, thereby avoiding any additional cost during inference.

For objective function, we use the combination of L1 and SSIM:
\begin{equation}
Loss=\frac{1}{T} \sum_{t=1}^T \alpha \Vert v^{'}_t-v_t \Vert_1 + (1-\alpha)(1-SSIM(v^{'}_t,v_t))
\label{13}
\end{equation}
where T is the frame number, $v^{'}_t$ and $v_t$ denote the reconstructed frame and its corresponding ground truth.

\section{Experiments}
\subsection{Settings}
\textbf{Datasets and metrics} Quantitative and qualitative comparison experiments are conducted on 8 different sequences of Big Buck Bunny~(Bunny) and UVG~\cite{mercat2020uvg} dataset.
Bunny has 132 frames with the resolution of 720$\times$1280.
The UVG dataset consists of 7 videos with the resolution of 1080$\times$1920, six of which contain 600 frames, and the remaining one has 300 frames.
For evaluation metrics, we use PSNR and MS-SSIM~\cite{wang2003ms-ssim} to evaluate the video reconstruction quality, where higher values indicate better quality.
Bits per pixel~(bpp) is used to measure the bitrate in video compression.
Model complexity is assessed based on model size, floating point operations~(FLOPs), training speed and decoding FPS.

\noindent\textbf{Implementation details} We utilize the Adam optimizer~\cite{kingma2015adam} with (0.5, 0.999) betas. The learning rate is 5e-4 with a cosine annealing schedule~\cite{loshchilov2022sgdr}.
For the loss in Eq.~(\ref{13}), the $\alpha$ is set to 0.7.
The upsampling factors for 720p and 1080p reconstruction are (5, 2, 2, 2, 2) and (5, 3, 2, 2, 2).
We further select three typical NVR methods as baselines, \ie E-NeRV~\cite{li2022e-nerv}, HNeRV~\cite{chen2023hnerv}, PNeRV~\cite{zhao2024pnerv}, and evaluate our proposed method on them.
Unless otherwise specified, we follow their original settings during training.
All experiments are implemented on an NVIDIA V100 GPU using PyTorch framework.

\begin{table}[t]
  \centering
  \caption{Average PSNR/MS-SSIM across eight videos under different training times, bold values indicate better performance.}
    \begin{tabular}{ccccc}
    \toprule
    Training time & 30min & 60min & 90min & 120min \\
    \midrule
    NeRV  & 28.94/0.8541 & 30.36/0.8825 & 31.09/0.8959 & 31.53/0.9034 \\
    Online-RepNeRV & \textbf{29.93/0.8697} & \textbf{31.07/0.8956} & \textbf{31.58/0.9043} & \textbf{31.87/0.9086} \\
    \midrule
    E-NeRV & 30.31/0.8733 & 32.05/0.9038 & 32.87/0.9175 & 33.47/0.9255 \\
    Online-RepENeRV & \textbf{30.43/0.8738} & \textbf{32.32/0.9079} & \textbf{33.24/0.9229} & \textbf{33.81/0.9299} \\
    \midrule
    HNeRV & 29.50/0.8671 & 30.32/0.8817 & 30.65/0.8868 & 30.79/0.8889 \\
    Online-RepHNeRV & \textbf{29.89/0.8730} & \textbf{30.57/0.8864} & \textbf{30.81/0.8900} & \textbf{30.98/0.8926} \\
    \midrule
    PNeRV & 27.57/0.8233 & 29.83/0.8700 & 31.54/0.8973 & 32.44/0.9084 \\
    Online-RepPNeRV & \textbf{30.23/0.8720} & \textbf{32.24/0.9022} & \textbf{33.09/0.9130} & \textbf{33.68/0.9216} \\
    \bottomrule
    \end{tabular}%
  \label{tab1}%
\end{table}%
\begin{table}[t]
  \centering
  \caption{PSNR/MS-SSIM on selected videos at the 30-minute training mark.}
  \resizebox{\textwidth}{!}{
    \begin{tabular}{ccccccc}
    \toprule
    Video & Bunny & Beauty & Bosphorus & Shakendry & Yachtride \\
    \midrule
    NeRV  & 31.65/0.9535 & 30.94/0.8618 & 30.33/0.8830 & 31.85/0.9176 & 25.10/0.7927 \\
    Online-RepNeRV & \textbf{33.34/0.9713} & \textbf{31.51/0.8722} & \textbf{31.14/0.8952} & \textbf{33.43/0.9335} & \textbf{25.79/0.8151} \\
    \midrule
    E-NeRV & 34.37/0.9804 & 31.69/0.8934 & 31.82/0.9059 & 34.24/0.9436 & 25.83/0.8174 \\
    Online-RepENeRV & \textbf{34.90/0.9827} & \textbf{31.70/0.8941} & \textbf{32.15/0.9119} & \textbf{34.40/0.9449} & 25.71/0.8123 \\
    \midrule
    HNeRV & 31.93/0.9648 & 31.71/0.8762 & 30.93/0.8897 & 32.57/0.9244 & 25.57/0.8167 \\
    Online-RepHNeRV & \textbf{32.43/0.9694} & \textbf{32.28/0.8838} & \textbf{31.16/0.8934} & \textbf{32.91/0.9271} & 25.46/0.8108 \\
    \midrule
    PNeRV & 30.48/0.9319 & 30.04/0.8455 & 28.69/0.8244 & 28.46/0.8178 & 24.28/0.7258 \\
    Online-RepPNeRV & \textbf{35.76/0.9802} & \textbf{31.00/0.8599} & \textbf{30.10/0.8612} & \textbf{32.30/0.9120} & \textbf{26.31/0.7892} \\
    \bottomrule
    \end{tabular}%
    }
  \label{tab2}%
\end{table}%

\subsection{Results}
As structural reparameterization expands the training-time architecture, making comparisons based on the same number of epochs may lead to unfair conclusions.
To ensure a fair evaluation, we instead adopt a fixed training time protocol, conducting experiments on NeRV, E-NeRV, HNeRV, and PNeRV under identical training durations of 30, 60, 90, and 120 minutes.

Table.~\ref{tab1} presents the average performance of various methods across eight datasets.
The results demonstrate that our proposed method consistently enhances the performance of four baseline models, confirming its effectiveness and versatility.
Notably, our methods achieve the most significant performance gains during the early stages of training.
For example, at the 30-minute mark, it yields PSNR improvements of 0.99 dB on NeRV, 0.39 dB on HNeRV, and a notable 2.66 dB on PNeRV. At 90 minutes, it improves E-NeRV by 0.37 dB.
This early performance gains highlights the advantage of online reparameterization in accelerating the convergence process.
In table.~\ref{tab2}, we show detailed results on a subset of videos, where our methods outperform the baselines on most cases, demonstrating its robust performance across diverse content.

\begin{figure}[t]
\includegraphics[width=\textwidth]{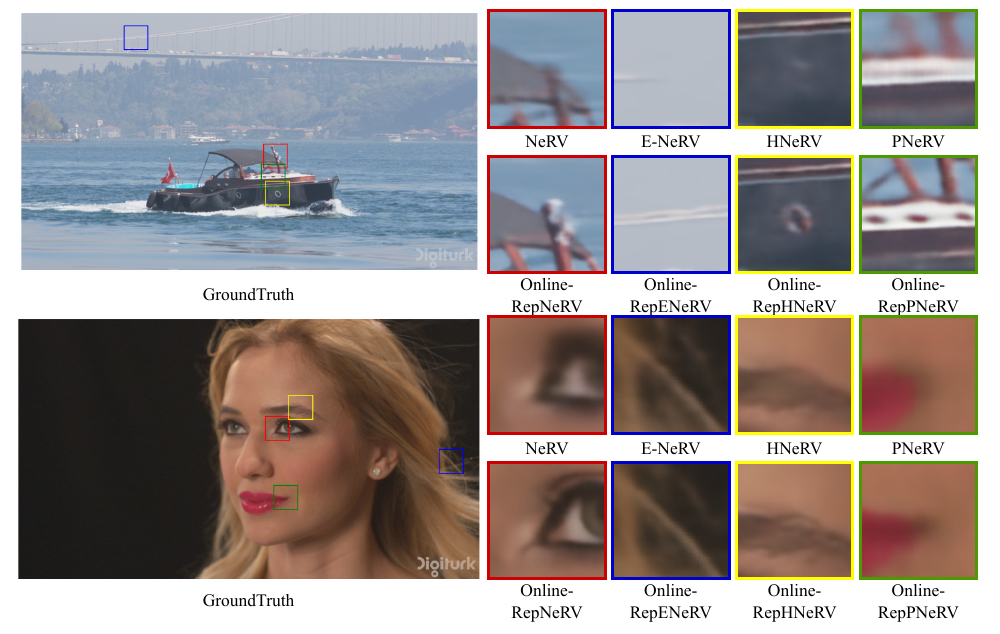}
\caption{Visual results for Bosphorus and Beauty.} \label{fig4}
\end{figure}

We also show qualitative improvements in Fig.~\ref{fig4}.
Compared to baselines, the reparameterized models exhibit more reliable reconstruction, particularly in regions where baselines struggle to recover.
Moreover, our method can produce sharper textures and improved visual fidelity.
These visual enhancements demonstrate the effectiveness of our method in improving both perceptual quality and reconstruction accuracy.

\begin{figure}[t]
\centering
\begin{minipage}{0.49\textwidth}
\includegraphics[width=0.9\linewidth]{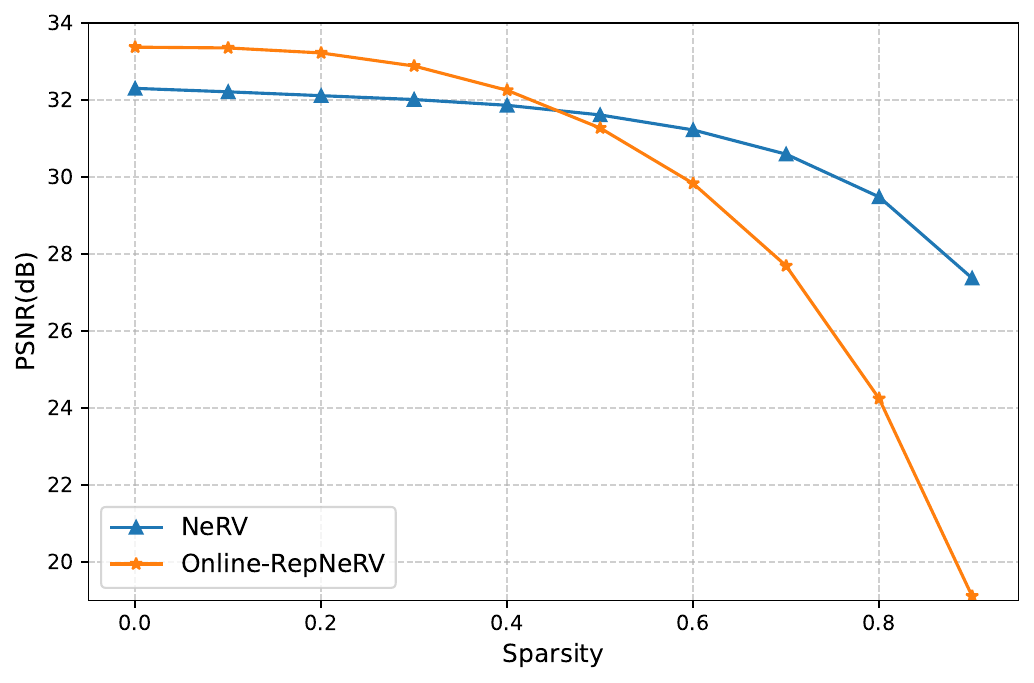}
\captionsetup{width=0.9\linewidth}
\caption{Pruning. Sparsity is the ratio of parameters pruned.}
\label{fig.5}
\end{minipage}
\begin{minipage}{0.49\textwidth}
\includegraphics[width=0.9\linewidth]{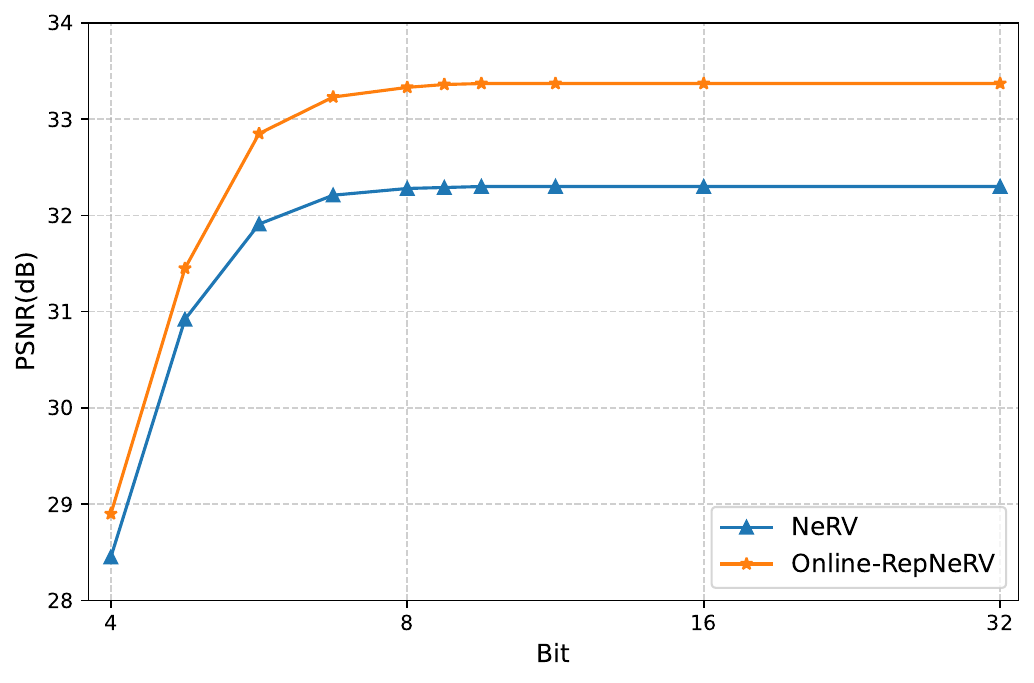}
\captionsetup{width=0.9\linewidth}
\caption{Quantization. Bit is the storage precision of the parameters.}
\label{fig.6}
\end{minipage}
\end{figure}

\begin{figure}[t]
\includegraphics[width=\textwidth]{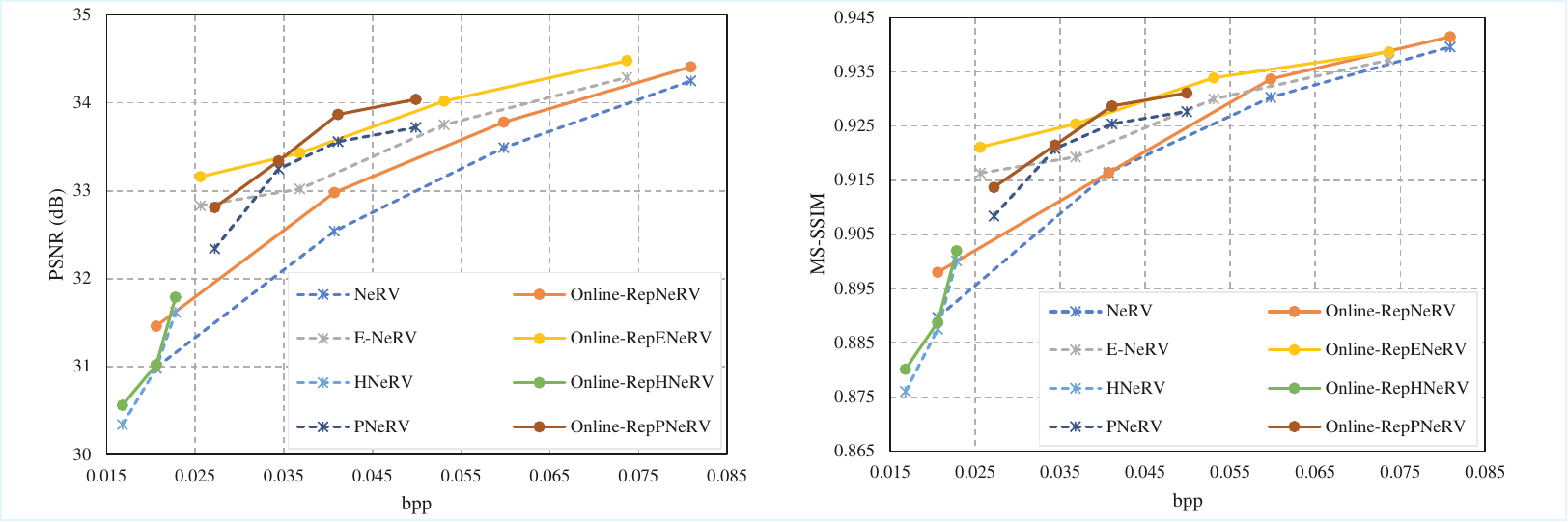}
\caption{Compression results on UVG dataset.} \label{fig7}
\end{figure}

For video compression task, we first perform ablation experiments on NeRV.
Fig.~\ref{fig.5} and Fig.~\ref{fig.6} illustrate the model pruning and quantization curves on Bunny, respectively.
For pruning, Rep-NeRV maintains superior performance at low sparsity levels, however, as the sparsity increases beyond 40\%, its performance begins to decline sharply.
Compared to NeRV, Rep-NeRV exhibits greater sensitivity to parameter reduction, which highlights an inherent limitation of structural reparameterization techniques.
The model benefits from a multi-branch structure during training, which introduces additional capacity and structural diversity.
These branches are fused into a single set of parameters for inference, resulting in a compact and tightly coupled representation.
Consequently, the merged structure lacks the redundancy typically present in overparameterized models, making it vulnerable to aggressive pruning.
For quantization, Rep-NeRV exhibits robustness in the full-parameter setting.
With the precision gradually decreasing, Rep-NeRV follows the same trend as NeRV while consistently achieving better results across all bit-widths.

We then conduct video compression experiments on NeRV, E-NeRV, HNeRV, and PNeRV.
With pruning~(10\% pruned), quantization~(8 bits), and entropy encoding, Fig.~\ref{fig7} shows the rate-distortion curves on UVG dataset.
Compared to baselines, our method yields more favorable performance curves, which indicates better rate-distortion trade-offs.

\subsection{Ablation Study}
\label{Ablation Study}
\textbf{Block structure} To validate the design of the proposed ECB, we conduct a series of ablation experiments on the constituent branches derived from the classical reparameterization blocks mentioned in Section \ref{Motivation}.
The results are shown in Table.~\ref{tab3}.
Among the most effective branches and filter combinations(\ie 3$\times$3, 1$\times$3, 3$\times$1, 1$\times$1-3$\times$3, and Sobel), we replace the 1$\times$1-3$\times$3 branch with 1$\times$1-3$\times$3-1$\times$1 to further enhance representational capacity.
Under this configuration, empirical results indicate that the inclusion of the Sobel filter yields negligible contributions to performance.
Consequently, we remove it from the final design.
The resulting ERB consistently outperforms alternative branch combinations, demonstrating its effectiveness.

\noindent\textbf{Model complexity} To evaluate the impact of structural reparameterization and the online reparameterization strategy, we conduct a comparative analysis of model complexity across NeRV, Rep-NeRV, and Online-RepNeRV.
As summarized in Table.~\ref{tab4}, Rep-NeRV introduces more parameters during training, its computational cost also increases proportionally with the number of expanded branches, consequently resulting in slower training speed.
However, Online-RepNeRV effectively mitigates the training overhead and accelerates the training speed to a level comparable with that of NeRV.
Notably, both Rep-NeRV and Online-RepNeRV maintain the same lightweight architecture as NeRV during inference, thereby incurring no additional runtime cost.

\begin{table}[t]
  \centering
  \caption{Ablation study on the components of the reparameterization block.}
    \resizebox{\textwidth}{!}{
    \begin{tabular}{ccrrrcc}
    \toprule
    3×3   & 1×3\&3×1 & \multicolumn{1}{c}{1×1} & \multicolumn{1}{c}{1×1-3×3} & \multicolumn{1}{c}{Scaled Filter} & 1×1-3×3-1×1 & PSNR/MS-SSIM \\
    \midrule
    \Checkmark     &       &       &       &       &       & 32.11/0.9587 \\
    \Checkmark     & \Checkmark     &       &       &       &       & 32.94/0.9666 \\
    \Checkmark     &       & \multicolumn{1}{c}{\Checkmark} &       &       &       & 32.56/0.9627 \\
    \Checkmark     & \Checkmark     & \multicolumn{1}{c}{\Checkmark} &       &       &       & 32.92/0.9670 \\
    \Checkmark     &       &       & \multicolumn{1}{c}{\Checkmark} &       &       & 33.05/0.9693 \\
    \Checkmark     & \Checkmark     &       & \multicolumn{1}{c}{\Checkmark} &       &       & 33.34/0.9705 \\
    \Checkmark     &       & \multicolumn{1}{c}{\Checkmark} & \multicolumn{1}{c}{\Checkmark} &       &       & 33.26/0.9702 \\
    \Checkmark     & \Checkmark     & \multicolumn{1}{c}{\Checkmark} & \multicolumn{1}{c}{\Checkmark} &       &       & 33.23/0.9702 \\
    \Checkmark     & \Checkmark     &       & \multicolumn{1}{c}{\Checkmark} & \multicolumn{1}{c}{AvgPool} &       & 33.36/0.9709 \\
    \Checkmark     & \Checkmark     &       & \multicolumn{1}{c}{\Checkmark} & \multicolumn{1}{c}{Sobel\&Laplacian} &       & 33.36/0.9709 \\
    \Checkmark     & \Checkmark     &       & \multicolumn{1}{c}{\Checkmark} & \multicolumn{1}{c}{Laplacian} &       & 33.23/0.9703 \\
    \Checkmark     & \Checkmark     &       & \multicolumn{1}{c}{\Checkmark} & \multicolumn{1}{c}{Sobel} &       & 33.37/0.9709 \\
    \Checkmark     & \Checkmark     &       &       & \multicolumn{1}{c}{Sobel} & \Checkmark     & 33.57/0.9723 \\
    \Checkmark     & \Checkmark     &       &       &       & \Checkmark     & \textbf{33.57/0.9727} \\
    \bottomrule
    \end{tabular}%
    }
  \label{tab3}%
\end{table}%

\begin{table}[t]
  \centering
  \caption{Comparison of model complexity on Bunny. Params and FLOPs are presented for training/inference, training speed means time(s)/epoch.}
    \begin{tabular}{lcccc}
    \toprule
    Model & Params(M) & FLOPs(G) & Training speed & Decoding FPS \\
    \midrule
    NeRV  & 3.20/3.20 & 100.94/100.94 & 7.61 & 75.34 \\
    Rep-NeRV & 7.58/3.20 & 420.17/100.94 & 28.45 & 77.37 \\
    Online-RepNeRV & 7.58/3.20 & 101.66/100.94 & 8.24 & 77.07 \\
    \bottomrule
    \end{tabular}%
  \label{tab4}%
\end{table}%

\section{Conclusion}
In this paper, we propose Online-RepNeRV, a plug-and-play reparameterized framework designed to improve Neural Video Representation (NVR). 
Specifically, we present the first systematic exploration of structural reparameterization within NVR by introducing an Enhanced Reparameterization Block (ERB). 
ERB integrates multi-scale convolutional branches to significantly enrich feature representations during training.
After training, these branches can be equivalently fused into a single compact convolution, ensuring efficient inference without additional overhead.
To further improve training efficiency, we propose an online reparameterization strategy to dynamically merge branch parameters within ERB, substantially reducing training overhead and accelerating convergence. 
Extensive experiments validate that Online-RepNeRV can be flexibly integrated into various NVR architectures and consistently outperforms existing baselines, demonstrating its effectiveness and broad applicability.
This work provides a new perspective on designing inference-efficient and capacity-aware neural video representations.

\textbf{Acknowledgments.} This work is supported by the National Natural Science Foundation of China (Grant No. 62031013) and the Guangdong Province Key Construction Discipline Scientific Research Capacity Improvement Project (Grant No. 2022ZDJS117)

\bibliographystyle{splncs04}
\bibliography{references}

\end{document}